\documentclass[12pt]{article}
\usepackage{graphics}

\newcommand{\be}{\begin{equation}}
\newcommand{\ee}{\end{equation}}
\newcommand{\bea}{\begin{eqnarray}}
\newcommand{\eea}{\end{eqnarray}}
\newcommand{\ba}{\begin{array}}
\newcommand{\ea}{\end{array}}
\begin{document}
%
\vspace*{1.0cm}

\begin{center}
\baselineskip 20pt
{\Large\bf
An extended study on the supersymmetric SO(10) models with natural doublet-triplet splitting
}
\vspace{1cm}

{\large
Qian Wan \footnote{ E-mail: wanqian@pku.edu.cn}
and
Da-Xin Zhang\footnote{ E-mail: dxzhang@pku.edu.cn}
}
\vspace{.5cm}

{\baselineskip 20pt \it
School of Physics,
Peking University, Beijing 100871, China}

\vspace{.5cm}

\vspace{1.5cm} {\bf Abstract}
\end{center}
In the supersymmetric SO(10) models,
the doublet-triplet splitting problem can be solved through the Dimopoulos-Wilczek mechanism.
This mechanism is extended in the non-renormalizable version.
Improvement on the realistic model is also made.

\thispagestyle{empty}

\newpage

\section{Introduction}

The Supersymmetric (SUSY) Grand Unified Theories (GUTs) are very important to search for
physics beyond the Standard Model (SM).
Among these models, the SUSY SO(10) models\cite{clark,moha}
are very predictive\cite{ss1,ss2,ss3,ss4,ss5,ss6,ss7}.
In SUSY SO(10) models,
the fermions of the three generation are contained in three spinor representations ${16}$ ($\psi$),
including right-handed neutrinos which are responsible for explaining the neutrino oscillation data
through the seesaw mechanism.

The renormalizable SUSY SO(10) models can be constructed successfully to achieve many
important features in the same models, such as gauge coupling unification, fermion masses, proton decays,
and doublet-triplet splitting (DTS).
In contrast,
the non-renormalizable (NR) SUSY SO(10) models are more difficult to be constructed successfully,
since in these models we need to take into account all allowed interactions which are consistent with
the symmetry of the models, even these interactions come from higher order operators.
These difficulties are very transparent in the realization of the DTS
through the Dimopoulos-Wilczek (DW) mechanism\cite{dw,dw0}.
The only known DW mechanism in the NR SUSY SO(10) models was realized in \cite{barrraby},
in contrast to those in the  renormalizable models \cite{dw0,dw3,dw4,dw10}.
The DW mechanism in the NR SUSY SO(10) models was further studied in \cite{babu2011} in
an effort to construct a realistic model.
However, to construct a successful fermion mass sector,
the model in \cite{babu2011} is very complicated and even incomplete.

In the present work,
we will extend the DW mechanism in the NR models and
improve the model in \cite{babu2011}.
We will first give an extended study on the NR version of the
DW mechanism.
We find that the DW mechanism can be applied in the presence of
new  high dimension couplings.
Then by modifying the model in \cite{babu2011},
we solve the difficulties in the fermion mass sector.
We will also discuss the proton decay problem in the new model.

\section{Extended study on the DW mechanism}

In SO(10) models, symmetry breaking requires the existence of two sectors in general.
The first sector, which breaks SO(10) into its subgroup
$SU(3)_c\times SU(2)_L \times SU(2)_R\times U(1)_{B-L}$ or $SU(3)_c\times SU(2)_L \times U(1)_{I_{3R}}\times U(1)_{B-L}$,
needs Higgs fields in the SO(10) representation  $45$ or $210$.
The second sector  is  rank breaking which breaks $SU(2)_R\times U(1)_{B-L}$ or
$U(1)_{I_{3R}}\times U(1)_{B-L}$ into $U(1)_Y$ of the SM.
This rank breaking sector needs fields carrying $B-L$ numbers to develop
Vacuum Expectation Values (VEVs).
The required fields are usually $126+\overline{126}$ in the renormalizable models
or $16+\overline{16}$ in the NR models.
These two sectors need to be coupled together to avoid extra massless Goldstones
which are not permitted at low energy.

The problem of DTS in GUT theories is to answer the question
why  the weak doublets of the minimal SUSY Model (MSSM) are so light
compared to the color triplets in the same Higgs multiplets.
In realizing DTS through the DW mechanism,
there are several versions can be used.
In the simplest NR models,  a $45$-plet ($A$) is needed with the
superpotential
\be
W_{DW}=\frac{m_A}{2} A^2+\frac{1}{M_*}A^4+\cdots,
\label{DW1}
\ee
where $M_*$ stands for some large mass scale,
and terms of higher dimensions are not displayed  explicitly.
The symbol $A^4$ are SO(10) invariants of the form
$(A^2)_X(A^2)_X$, where $X=1,54,210,770$\cite{slansky} are symmetric representations of SO(10),
so the second term in (\ref{DW1}) stands for four terms with
different couplings which
are  suppressed  for simplicity.
In the following, analogue couplings will also be suppressed.

After symmetry breaking, the field $A$ can develop two VEVs  along the SM singlet
directions.
They are labeled by the representations under the SO(10) subgroup
$SU(4)_c\times SU(2)_L \times SU(2)_R$
as $A_1(1,1,3)$ and $A_2(15,1,1)$.
The VEV in (\ref{DW1}) is in the form
\be
\langle W_{DW}\rangle=\frac{m_A}{2} (A_1^2+A_2^2)+\frac{1}{M_*}(A_1^4+A_1^2A_2^2+A_2^4).
\label{DW2}
\ee
In the bracket of the second term, we have also suppressed the CG coefficients.
In (\ref{DW2}) the dependence on $A_1$ is at least quadratic and thus
the F-term corresponding to $A_1$ is proportional to $A_1$.
Then, the F-flatness conditions give several solutions, among them we pick up the DW solutions
\be
A_1=0, ~A_2\sim \sqrt{m_A M_*}
\label{solution}
\ee
to realize DTS when coupled with two different  Higgs in the $10$\cite{dw}.
In additional to (\ref{DW2}), the higher  dimensional terms with even numbers of $A$s
are also of the forms $A_1^{2m}A_2^{2n}$ ($m,n$ are integers).
There are also higher  dimensional terms with odd numbers of $A$s in the form
\be
\epsilon \cdot A^5\equiv \epsilon^{i_1i_2\cdots i_{10}}A_{i_1i_2}\cdots A_{i_9i_{10}},
\label{A5}
\ee
whose contribution to the VEV is $A_1^2 A_2^3$.
All these higher terms do not spoil the DW solutions (\ref{solution}).

The difficulty in the NR models is that the fields
$A$(45) need to be coupled with the rank breaking Higgs fields
$16+\overline{16}$.
These couplings include those which are linear in $A_1$.
When the linearly coupled terms are added in additional to (\ref{DW2}),
the DW solutions disappear.
In \cite{barrraby} the difficulty is overcome by introducing two pairs of
$16+\overline{16}$ ($C,\bar{C},C',\bar{C}'$) through
\begin{equation}
W_{Rank}=\left(\bar{C}C'+\bar{C}'C\right)
\left(S+\frac{1}{M_*}\bar{C}C+\frac{1}{M_*}ZA\right)+\frac{1}{M_*^2}\bar{C}'C'Z^2S,
\label{barr}
\end{equation}
and pick up the solutions $\langle C^\prime\rangle=\langle \overline{C^\prime}\rangle=0$.
Here $S$ and $Z$ are SO(10) singlets.
Later on, in \cite{babu2011} a realistic model has been attempted to construct.

Together with (\ref{DW1}) and (\ref{barr}),
the SO(10) symmetry is broken down into the SM gauge symmetry.
The F- and D-flatness conditions give
\begin{equation}
s\sim \frac{c^2}{M_*},~ z\sim \frac{c^2}{A_2},~ A_2\sim \sqrt{m_AM_*},
\end{equation}
where $s=\langle S\rangle$, $z=\langle Z\rangle$ and $c=\langle C\rangle\equiv \langle \bar{C}\rangle$.
As will be introduced in the next section, an anomalous gauge U(1) symmetry
is usually used to forbid the unwanted couplings of a model.
The D-term of this symmetry has a
Fayet-Iliopoulos term  of the order $-\sqrt{\xi}\sim (0.1-1) M_*\sim 10^{17-18}$GeV \cite{green,bere},
which is to be canceled by the VEVs of those
 U(1) charged fields $\phi_i$ as $\xi+\sum Q_i \phi_i^2=0$.
Typically we have $c^2+s^2+A_2^2\sim -\frac{1}{4}\xi$ \cite{babu2011}, then
\begin{equation}
	c,z\sim (1-10)\times M_{GUT}\,;\quad s\sim (0.01-0.1)\times M_{GUT}\,;\quad A_2\approx M_{GUT}.
\end{equation}
These VEVs will be used in the following study on the fermion sector of the MSSM.

To generate the massless doublets of the MSSM while keeping the color triplets heavy,
two $10$-plets $H$ and $H'$ are introduced to couple with $A(45)$ with the DW solutions through
\be
HAH'+\frac{m_{H'}}{2}H'H'.
\label{H}
\ee
Here $H=H(1,2,2)+H(6,1,1)$ (and similar for $H'$).
This is the standard form of superpotential in realizing DTS through the DW mechanism\cite{dw}.
In the NR models,
there are also higher  dimensional terms in additional to (\ref{H})
in the form
\be
\frac{1}{M_*^{2m}}H A^{2m+1} H' ~(m={\rm integer})
\ee
which does not give mass to the doublets  $H(1,2,2)$ since $A_2^{2m+1}$
contain no VEV  in either  (1,1,1) or (1,1,3) direction.
Terms $H A^{2m} H'$ will spoil the DW mechanism which need to be forbidden
by imposing extra symmetries on the models.

In the present study we need to extend the coupling $HAH'$
to the case in  the presence of spinors  $C(16), \bar{C}(\overline{16})$.
These couplings are of the forms
\begin{equation}
\frac{1}{M_*}HA\left(C^2+\bar{C}^2\right),
\label{C1}
\end{equation}
which have escaped  attentions  in the literature.
Note that $16\times 16=10_S+120_A+126_S, \overline{16}\times \overline{16}=10_S+120_A+\overline{126}_S$,
where the labels $A,S$ stand for anti-symmetric and symmetric, respectively,
$CC$ and $\bar{C}\bar{C}$ thus contain no $120$.
Also, because $10,126+\overline{126},45$ are tensors of ranks 1,5,2, respectively,
they do not couple together.
Consequently, in (\ref{C1})
$CC$ and $\bar{C}\bar{C}$ can be only in the representation $10$
so that the DW mechanism can be applied in this case.

The operators of the form $HA^{2m}(C^2+\overline{C}^2)$ also
destroy the DW mechanism which need to be forbidden by introducing an extra symmetry,
as in the case of the operators $HA^{2m}H^\prime$.
In the presence of the higher  dimensional couplings
\be
\frac{1}{M_*^{2m+1}}HA^{2m+1}\left(C^2+\bar{C}^2\right),
\label{C2}
\ee
$CC$ ($\bar{C}\bar{C}$) can be either $10$ or $126$ ($\overline{126}$).
The case $(CC)_{126}$ (and similarly $(\bar{C}\bar{C})_{\overline{126}}$) needs to be discussed
as following.
Note that $10\times 126=210+1050$\cite{slansky},
neither $210$ nor $1050$ has a SM singlet direction which couples with odd number of $A_2$ fields.
More generally, two fields
(or two products of  fields with each product in a  irreducible representation)
can only couple to
VEVs with either even or odd numbers of $A_2$ fields.
The exceptions come from the couplings of the form (\ref{A5}) which are zero in the  DW cases with $A_1=0$.
This completes the feasibility of  the DW mechanism in the presence of the higher  dimensional couplings
(\ref{C2}).

\section{Toward a realistic model}

The model in  \cite{babu2011} has difficulties in the SM
fermion mass sector, where a complete set of Yukawa couplings were not given.
In the present work,
we will improve the model in \cite{babu2011} to solve the difficulties.
As in \cite{babu2011},
the Higgs sector contains an adjoint  $A(45)$, two $10$-plets $H+H'$,
two pairs of spinor-anti-spinor superfields $C/\bar{C}+C'/\bar{C}'$ and two singlets  $S$ and $Z$.
A $Z_2$ assisted anomalous $U(1)$ symmetry is needed to discard
harmful couplings.
The charges of the Higgs fields and those of the three matter families $f_i$ under $U(1)_A\times Z_2$ are listed in Table \ref{tab}.

\begin{table*}[h]
	\centering
	\begin{tabular}{c|cccc|ccc|cc|c}
		& $\bar{C}$ & $\bar{C}'$ & $C$ & $C'$ & $S$ & $Z$ & $A$ & $H$ & $H'$ & $f_i$ \\ \hline
		Q & 1 & -7 & 3 & -5 & 4 & 4& 0 & 6 & -6 & -3 \\
		$Z_2$ & + & + & + & + & + & - & - & + & - & + \\
	\end{tabular}
	\label{tab}
	\caption{$U(1)_A\times Z_2$ charges for all the superfields.}
\end{table*}

The main difference of the charges in Table \ref{tab}  from those in \cite{babu2011} are
introduced to replace the Higgs coupling $H\bar{C}\bar{C}$ by $H\bar{C}\bar{C}'$.
The superpotential for the Higgs sector is
\begin{eqnarray} W_H&=&HH'A+H\bar{C}\bar{C}'+\frac{1}{M_*}(H'\bar{C}^2Z+H'AC^2+HC'^2S)+\frac{S}{M_*^2}(H'CC'Z+H'^2Z^2+H'A\bar{C}^2) \nonumber\\
		 &+&\frac{Z^2}{M_*^3}(H'\bar{C}\bar{C}'Z+H'^2C\bar{C}+H'^2AZ+H'ACC'+H'A\bar{C}^3C)
+\cdots,
\end{eqnarray}
following which the electro-weak doublets have the mass matrix
\begin{equation}
	M_D=\left(\begin{array}{cccc}
		0 & 0 & c & 0 \\
		0 & 0 & s+\frac{c^2}{M_*}+\frac{zA_2}{M_*} & 0 \\
		0 & s+\frac{c^2}{M_*}+\frac{zA_2}{M_*} & \frac{sz^2}{M_*^2}
& \frac{csz}{M_*^2}+\frac{cz^2A_2}{M_*^3} \\
		0 & \frac{cz}{M_*}+\frac{csA_2}{M_*^2} &
\frac{cz^3}{M_*^3}+\frac{cz^2sA_2}{M_*^4} & M_{H'}
	\end{array}\right)
\label{MD}
\end{equation}
where the columns are $H_u,\bar{C}_u,\bar{C}_u',H_u'$ and the rows are $H_d,C_d,C_d',H_d'$, respectively,
and
\begin{equation}
	M_{H'}=\frac{c^2+zA_2}{M_*}+\frac{sA_2^2+sz^2}{M_*^2}+\cdots.
\nonumber
\end{equation}
In the 4th column, the entry (1,4) comes from the coupling $HAH'$ so that only the VEV $A_1=0$ can be taken.
The entry (2,4) comes from $H'ACC$ where the product $CC$ can only take the representation $10$ of SO(10),
and again, only the VEV $A_1=0$ can be  taken.
As was discussed after (\ref{C2}), higher dimensional operators do not break this result.

The massless eigenstates of (\ref{MD}) are
\begin{equation}
H_u^0=H_u,~~~ H_d^0=\frac{c}{\sqrt{c^2+M_*^2}}H_d-\frac{M_*}{\sqrt{c^2+M_*^2}}C_d,
\label{H0}
\end{equation}
they are the Higgs doublets of the MSSM.
In contrast, in \cite{babu2011} the $H_d^0$ has components from all the four down-type Higgs doublets,
which differ from the present model in the fermion sector of the MSSM.

In the spectrum, there is no major difference between this work
and \cite{babu2011}. As in \cite{babu2011}, threshold effects
can be adjusted to be small
 in realizing gauge coupling unification. We refer \cite{babu2011} for more details.

\section{Fermion masses and proton decays}

With Table \ref{tab}, the most general Yukawa couplings are
\begin{eqnarray}
		W_Y&=& f_if_j\left(H+\frac{1}{M_*}C^2+\frac{1}{M_*^2}(\bar{C}^2S+HA^2)+\frac{Z^2}{M_*^3}(CC'+H'Z)
+\frac{SZ^2}{M_*^4}\bar{C}\bar{C}'\right. \nonumber\\
		&+&\left.\frac{Z^4}{M_*^5}C'^2+\frac{SZ^4}{M_*^6}\bar{C}'^2\right),
\label{yukawa}
\end{eqnarray}
in which the fermion masses are given by
\begin{equation}
f_if_j\left(H+\frac{C^2}{M_*}+\frac{\bar{C}^2S}{M_*^2}+\frac{HA^2}{M_*^2}\right)\supset f_if_j\left((H_u+H_d)+2\frac{c}{M_*}C_d+\frac{s}{M_*^2}\bar{C}^2+\frac{A_2^2}{M_*^2}(H_u+H_d)\right),
\label{y1}
\end{equation}
where the third term  contributes to the Majorana masses for the right-handed neutrinos
at $sc^2/M_*^2 \sim 10^{14}\,\mathrm{GeV}$ when $s\sim 0.1M_{GUT}$ is taken.
This term contributes neither to Dirac masses for the neutrinos nor to the charged fermion masses,
so data on large  neutrino mixing can be fitted without further constraints.
In the  fourth term, $HA^2$ can be $10$ or $\overline{126}$ but not $120$
when $A_1=0$ is taken, and $\overline{126}$
gives Georgi-Jarlskog type corrections to  fermion masses\cite{georgiJ}.
(\ref{y1}) also contribute to proton decays through dimensional-five operators
when replacing the doublets by
the color triplets of the same representations.

In (\ref{yukawa}),
\begin{equation}
f_if_j\left(\frac{CC'Z^2}{M_*^3}+\frac{H'Z^3}{M_*^3}+\frac{\bar{C}\bar{C}'SZ^2}{M_*^4}\right)
\nonumber
\end{equation}
contributes to proton decays but not to fermion masses,
but these contributions are power suppressed so the couplings can be set to zeros safely.
The remaining terms in (\ref{yukawa}),
\begin{equation}
f_if_j\left(\frac{C'^2Z^4}{M_*^5}+\frac{\bar{C}'^2SZ^4}{M_*^6}\right)
\nonumber
\end{equation}
contribute to neither, thus are irrelevant to the present study.

In summary,
the Yukawa couplings (\ref{yukawa}) can be simplified as
\begin{equation}
f_if_j\left(Y_H H+Y_C\frac{1}{M_*}(C^2)_{10}+Y_{\bar{C}}\frac{s}{M_*^2}(\bar{C}^2)_{\overline{126}}
+Y_{HAA}\frac{1}{M_*^2}(HA^2)_{\overline{126}}\right)_{ij},
\label{y4}
\end{equation}
where the Yukawa couplings $Y$'s are added explicitly.
We have been omitted the contributions from $(HA^2)_{10}$
which is a small correction to the $Y_H$ term.
Together with (\ref{H0}), the Dirac mass matrices of fermions are given by
\begin{eqnarray}
		M_u&=&\left(Y_H+Y_{HAA}\frac{A_2^2}{M_*^2}\right) \left<H_u^0\right>, \nonumber\\
		M_\nu&=&\left(Y_H-3Y_{HAA}\frac{A_2^2}{M_*^2}\right) \left<H_u^0\right>, \nonumber\\
		M_d&=&\frac{c}{M_*}\left(Y_H+2Y_C+Y_{HAA}\frac{A_2^2}{M_*^2}
		\right)\left<H_d^0\right>, \nonumber\\
		 M_e&=&\frac{c}{M_*}\left(Y_H+2Y_C-3Y_{HAA}\frac{A_2^2}{M_*^2}\right)\left<H_d^0\right>,\label{mferm}
\end{eqnarray}
following which a small $\tan \beta\equiv \frac{\left<H_u^0\right>}{\left<H_d^0\right>}\sim O(1)$ is preferred.
It can be also noted that in the down-type quark and charged lepton masses,
the $Y_{HAA}$ term is suppressed by a factor $\frac{A_2^2}{M_*^2}$
compared to the $Y_H$ term,
implying that the Georgi-Jarskog mechanism affects the first two generations  mainly.
Then at GUT scale we have $m_s\sim m_\mu$ instead of $m_s\sim \frac{1}{3} m_\mu$ in [19],
while the approximate relation $\frac{m_e}{m_\mu}\sim \frac{1}{9} \frac{m_d}{m_s}$
can be satisfied by adjusting the $Y_{HAA}$ terms in (\ref{mferm}).

Provided that (\ref{y4}) is sufficient in giving fermion masses
and thus $Y$'s are determined,
we can estimate the proton decay rates.
In SUSY GUT models, the dominant contributions for
proton decay originate from dimension-5 operators mediated by the
color-triplet Higgs(-inos). These operators are then dressed by the
wino-sfermion loops. The most typical mode is
$p\to K^+\overline{\nu}$
with the current limit  $\tau >6.6\times 10^{33}$ years.
With most SUSY parameters $3 -100$TeV,
the color-triplet Higgs are required to be $\geq 10^{18-19}$GeV \cite{ellis}.
In the present case,
not all of the 4 pairs of color triplets can mediate proton decay.
The amplitudes are then proportional inversely to the effective triplet masses resulting from integrating out those triplets which do not couple to the
fermion superfields \cite{lizhang}.
The mass matrix for the color triplet Higgs is analogue to (\ref{MD}),
\begin{equation}
	M_{T}=\left(\begin{array}{cccc}
		0 & 0 & c & A_2 \\
		0 & 0 & s+\frac{c^2}{M_*}+\frac{zA_2}{M_*} & \frac{cA_2}{M_*} \\
		0 & s+\frac{c^2}{M_*}+\frac{zA_2}{M_*} & \frac{sz^2}{M_*^2}
& \frac{csz}{M_*^2}+\frac{cz^2A_2}{M_*^3} \\
		A_2 & \frac{cz}{M_*}+\frac{csA_2}{M_*^2} &
\frac{cz^3}{M_*^3}+\frac{cz^2sA_2}{M_*^4} & M_{H'},
	\end{array}\right)
\label{MT}
\end{equation}
with the columns are $H_T,\bar{C}_T, \bar{C}_T',H_T'$ and the rows are
$H_{\overline{T}},{C}_{\overline{T}}, {C}_{\overline{T}}',H_{\overline{T}}'$, respectively.
An effective triplet mass $M_{ij}^{eff}$ corresponds to the matrix element (i,j) after integrating out
all the other elements in the triplet mass matrix,
and  is given by\cite{lizhang}
\begin{equation}
{M_{ij}^{eff}}=(-1)^{i+j}\frac{Det(M)}{Det(M_{ji}^\star)},
\end{equation}
where  $Det(M_{ji}^\star)$ is the cofactor
corresponding to the determinant of
the matrix $M$ with the jth row and the ith column eliminated.

For proton decay through dimension-5 operators, the coefficients are
\begin{equation}
Y_HY_H\frac{1}{M_{11}^{eff}}
+\frac{cs}{M_*^2}Y_HY_{\bar{C}}\frac{1}{M_{12}^{eff}}
+\frac{cs}{M_*^2}Y_HY_C\frac{1}{M_{21}^{eff}}
+\frac{c^2s}{M_*^3}Y_CY_{\bar{C}}\frac{1}{M_{22}^{eff}}.
\label{ampl}
\end{equation}
Numerically,
\begin{equation}
M_T\sim\left(\begin{array}{cccc}
	0 & 0 & 10^{16} & 10^{16} \\
	0 & 0 & 10^{15} & 10^{15} \\
	0 & 10^{15} & 10^{12} & 10^{12} \\
	10^{16} & 10^{16} & 10^{13} & 10^{13}
	\end{array}\right)\nonumber
\end{equation}
if we take $c,z,A_2\sim M_{GUT}\,;\quad s\sim 0.1M_{GUT}$,
\be
M_{11}^{eff}, M_{12}^{eff}\sim 10^{19}GeV,~~~ M_{21}^{eff}, M_{22}^{eff}\sim 10^{18}GeV.
\ee
Together with the suppressed Yukawa couplings in (\ref{ampl}),
the proton decay amplitudes can be suppressed sufficiently.

\section{Summary}
We have extended the DW mechanism in the NR models to the more general forms.
A realistic model is modified into a new version.
The fermion masses and
the proton decay suppressions are also studied.

\end{document}